# Room-Temperature Multiferroic Skyrmions in LiNbO$_3$ with enhancement in electric-optical property

Yalong Yu[a], Bo Xiong[a], Siqi Wu[b], Yekai Ren[a], Nuo Chen[a], Qingjiao Mi[a], Zhaojie Zheng[a], Kangping Lou[a], Rui Wang[a], and Tao Chu[a,*]

[a] College of Information Science and Electronic Engineering, Zhejiang University, Hangzhou 310058, P. R. China

[b] School of Physics, Zhejiang University, Hangzhou 310058, P. R. China

[*] Corresponding Author: Tao Chu  chutao@zju.edu.cn

## Abstract

**LiNbO$_3$ (LN) is renowned for its exceptional ferroelectric properties, particularly its notable linear electro-optical (EO) effect, which is highly advantageous for various applications such as high-speed communication, optical computation, and quantum information processing. Compared to its ferroelectric properties, the magnetism of LN is not attractive enough due to its weak ferromagnetic nature. Theoretical studies suggest that LN may exhibit a novel magnetoelectric coupling via ferroelectrically-induced ferromagnetism. However, this mechanism has not yet been experimentally validated in any materials, presenting significant challenges for research. In this study, we provide the first experimental evidence supporting the mechanism of ferroelectrically-induced ferromagnetism in LN, including observations of the Dzyaloshinskii-Moriya interaction (DMI) and magnetoelectric coupling. Additionally, we have identified various multiferroic skyrmions, within which ferroelectric polarization signals are detectable. These signals can** **be influenced by the magnetic vortex structures, indicating a magnetoelectric coupling nature. Currently, they are the only multiferroic skyrmions that can keep stable at room temperature. Moreover, these magnetic textures significantly affect the ferroelectric properties, as demonstrated by an enhancement of the linear electro-optic effect of LN by over 200%. Given the novel magnetoelectric coupling mechanism, the potential of multiferroic skyrmions in spintronics and advanced data storage, and the extensive use of LN EO modulators, our research has significant implications for condensed matter physics, multiferroic materials, and optoelectronics.**

## Introduction

$LiNbO_3$ (LN) is an artificial ferroelectric material that has been extensively researched over the past decades due to its potential as an ideal platform for high-speed communication, quantum computing, and photonic accelerators[1-3]. It exhibits various electrical properties, including piezoelectricity, thermoelectricity, and the electro-optic (EO) effect[4]. Consequently, LN is utilized in a range of devices such as acousto-optic modulators, EO modulators, radio-frequency filters, and piezoelectric sensors[5-8]. Recently, photonic integrated circuits (PICs), particularly those based on thin-film lithium niobate (TFLN), have attracted significant attention due to their potential for high-speed communication (theoretically over terahertz frequencies), owing to the high intrinsic linear EO coefficient (the Pockels coefficient) of LN[9,10]. This property enables the operation of linear and high-speed modulators at low voltage levels. PICs based on TFLN waveguides are expected to outperform those based on silicon waveguides, where signal conversion is achieved through the plasma dispersion effect—a method that presents significant challenges at frequencies above 150 GHz due to inherent physical limitations in efficiency and speed.

In addition to its well-known ferroelectric properties, the magnetic characteristics of LN are crucial for a wide range of applications, including data storage, energy conversion, and signal processing[11-13]. The investigation of magnetoelectric coupling—the interaction between electrical properties and magnetism—could advance condensed-matter physics and facilitate the development of electric-field-driven magnetic storage systems. While current magnetoelectric coupling materials mainly involve magnetically-induced ferroelectricity (requiring non-collinear spin structures or exchange striction magnetic structures)[14-17], LN does not fit these mechanisms. However, a theoretical mechanism called ferroelectrically-induced ferromagnetism (FE-induced-FM) has been proposed[18,19], suggesting that LN might exhibit this behavior[20,21]. This mechanism posits that if the midpoint between magnetic moments is an inversion center, ferroelectric polarization can break inversion symmetry, leading to a non-collinear magnetic state with weak ferromagnetism and Dzyaloshinskii-Moriya interaction (DMI)[22,23]. This could make electric field-induced magnetization switching more feasible in bulk materials compared to magnetically-induced ferroelectricity. $FeTiO_3$, a candidate material, exhibits a $LiNbO_3$-type phase under 18 GPa and meets criteria for FE-induced-FM, with observed ferroelectricity and weak ferromagnetism below 120 K[24]. However, further evidence of DMI or magnetoelectric coupling is still needed.

In this study, we explore why LN may satisfy the symmetry required for FE-induced-FM, which is briefly noted in previous research[21]. We present compelling evidence for the presence of DMI in LN through the discovery of magnetic skyrmions[25], which exhibit unique properties including persistence at room temperature without external fields and the ability to induce ferroelectric polarization within magnetic vortices. The skyrmions in LN (SK-LNs) are the first room-temperature multiferroic skyrmions reported. Additionally, we identify various hopfion-like spin textures where the ferroelectric domain morphology is influenced by the magnetic vortex structure, demonstrating magnetoelectric coupling. The excitation of SK-LNs enhances the linear EO effect of LN by over 200 %. Consequently, we have developed an EO modulator on thin-film lithium niobate on insulator (LNOI) with an exceptional half-wave voltage of only 0.63 V·cm, setting a leading benchmark in current technology.

**Identification of magnetic skyrmions**

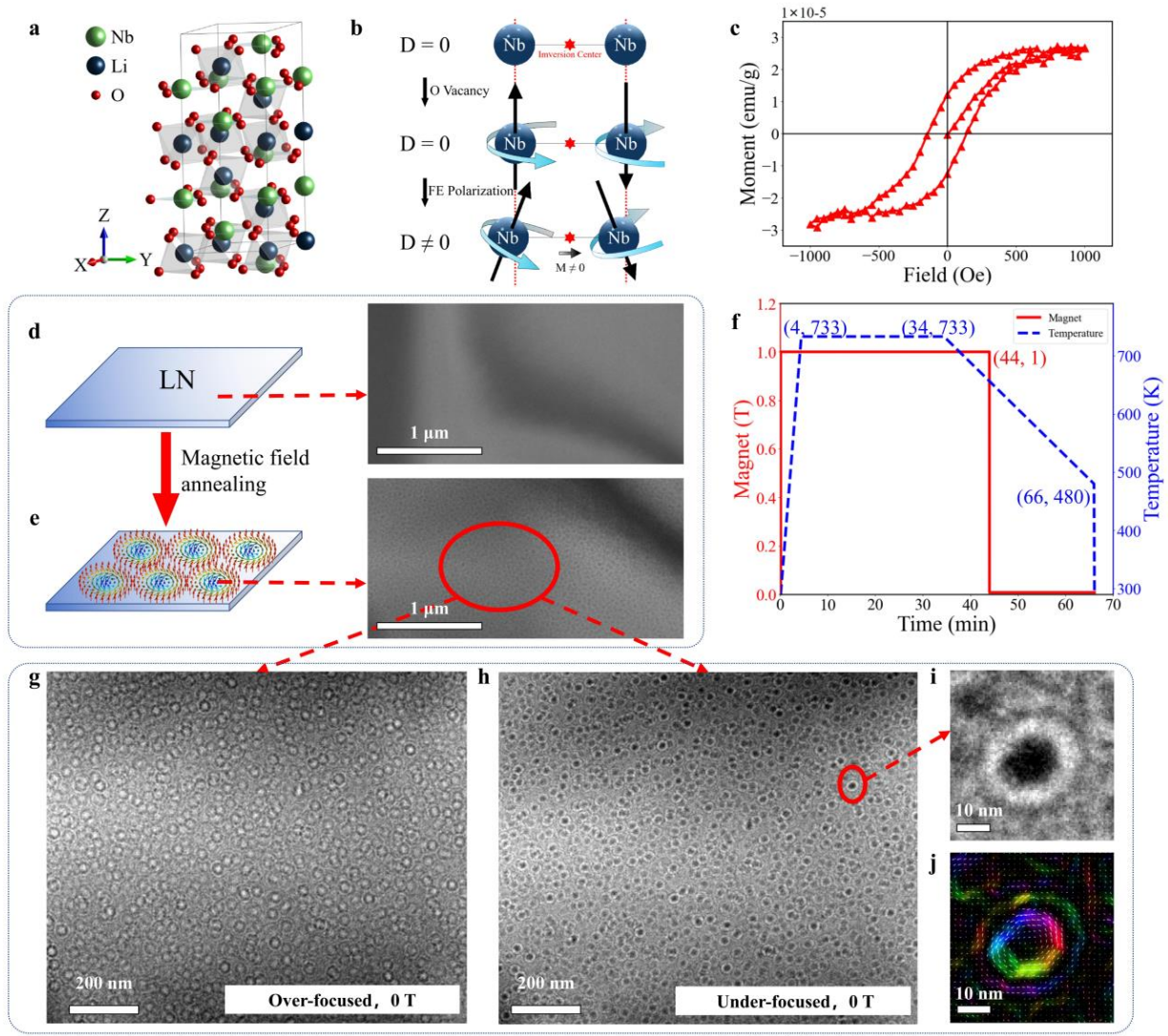

**Fig. 1** | **Room-temperature skymions in LN**. (**a**) Crystal structure of paraelectric LN. (**b**) Schematic diagram of FE-induced-FM in LN.The. (**c**) *M-H* curve of LN thin film measured at 300 K. (**d**,**e**) Under-focused LTEM images of LN before and after magnetic field annealing, respectively. (**f**) Process of magnetic field annealing. (**g**,**h**) Over- and under-focused LTEM images of LN after magnetic field annealing in a narrow field of view. (**i**) Single skyrmion shown in (**h**). (**j**) Magnetic induction map of (**i**), in which the colour and white arrows indicate the magnetisation direction at each point.

To understand why LN might satisfy the conditions for FE-induced-FM[18], consider the Hamiltonian for the DMI: $\vec{H}_{DM} = \vec{d_{ij}} \cdot (\vec{s_i} \times \vec{s_j})$, where $\vec{d_{ij}}$ is the Dzyaloshinskii vector, $\vec{s_i}$ and $\vec{s_j}$ are the localized magnetic moments at sites $i$ and $j$. When the midpoint between the magnetic sites $i, j$ coincide with the inversion center of the lattice, $|\vec{d_{ij}}|$ must be zero due to inversion symmetry. For the material whose paraelectric phase satisfing this condition, when it transforms into a ferroelectric phase below the Curie temperature, the ferroelectric polarization will disrupt the inverse symmetry of this magnetic property. In this case, the Hamiltonian can be expressed as: $H_{PLM} \sim \vec{P} \cdot (\vec{M} \times \vec{L})$, where $\vec{P}, \vec{M}, \vec{L}$ are the electric polarization, the magnetization, and the antiferromagnetic order parameter repectively. To minimize the system's energy, the interacting magnetic moments will undergo slight canting to induce a nonzero $\vec{M}$, resulting in non-collinear configurations and inducing DMI. For LN, the lattice structure in the paraelectric phase (as shown in Fig. 1(a)) has midpoints of Nd-Nd and Li-Li atoms aligning with the inversion center[26]. If antiferromagnetic coupling occurs between neighboring Nb atoms, LN could meet the criteria for FE-induced-FM. Density Functional Theory (DFT) calculations suggest that Nb-s electrons tend to align antiferromagnetically, particularly in the presence of oxygen vacancies[21], which are common during LN growth. This supports the possibility that LN may exhibit FE-induced-FM. The specific processes are elaborated in Fig. 1(b) to enhance readers' comprehension. The lattice structure of LN inherently shows paramagnetism, but defects in Li and Nb can induce ferromagnetism. This complicates the task of determining whether the observed weak ferromagnetism in LN is due to the FE-induced-FM mechanism. However, since FE-induced-FM is a unique source of DMI, magnetic skyrmions or helimagnetic strips emerging from DMI can serve as experimental evidence for this mechanism.

In this study, we utilized stoichiometric LN thin films a thickness of 500 microns. Magnetic measurements revealed a ferromagnetic nature in these films, although with weak intensity (see Fig. 1(c)). Fig. 1(d, e) show Lorentz transmission electron microscopy (LTEM) images of the LN thin film at 300 K without an external magnetic field (detailed in Extended Fig. 1). Fig. 1(d) presents under-focused LTEM images of the initial LN film, where no magnetic structures are observed, consistent with previous research. However, following magnetic field annealing (Fig.1(f)), numerous localized vortex patterns emerged, as depicted in Fig. 1(e). To further analyze these patterns, LTEM images of the LN films were examined under both over-focused and under-focused conditions with a focal-length deviation of 90 μm, using a narrower field of view, as shown in Figs. 1(g) and (h). The vortex patterns display opposite contrasts in under-focused and over-focused images, providing evidence of their magnetic skyrmion nature[27-30]. Fig. 1(i) and (j) show LTEM images of a single skyrmion in an under-focused state and its magnetic moment morphology, processed using the transport of intensity equation algorithm[31]. The skyrmion in LN has a typical size of 30 nm and a skyrmion number (Q) of 1. Notably, the Curie temperature of LN is 1483 K, which excludes the possibility of a structural phase transition. Thus, the presence of magnetic skyrmions in LN indicates the existence of DMI. This finding is significant as it provides the first substantial evidence for the existence of FE-induced-FM, surpassing the multiferroic characteristics observed in $FeTiO_3$.

## **Properties of SK-LNs**

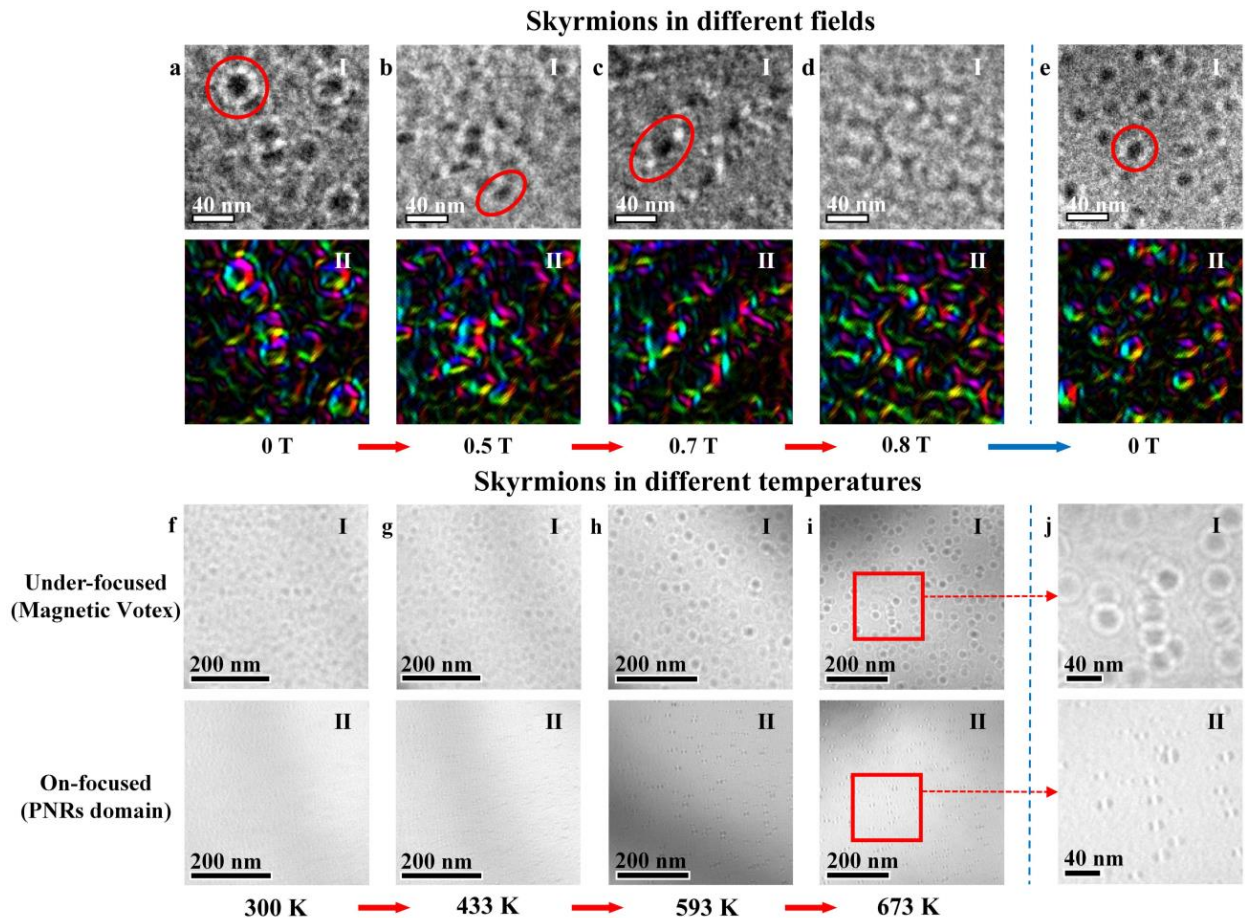

**Fig. 2 | Skyrmions under different conditions**. (I) Under-focused LTEM images showing skyrmions with applied magnetic fields of (**a**) 0 T, (**b**) 0.5 T, (**c**) 0.7 T, (**d**) 0.8 T, and (**e**) returning to 0 T, in which a typical skyrmion is marked by a red ring. (II) Corresponding magnetic induction maps. (I) Under-focused LTEM images showing skyrmions with applied temperatures of (**f**) 300 K, (**g**) 433 K, (**h**) 593 K, and (**i**) 673 K. (II) Corresponding on-focused LTEM images, in which the PNR texture can be observed. (**j**) Enlarged images of (**i**).

Given that DMI arises from the unique interplay between ferroelectric polarization and the ferroelectric environment of LN, SK-LNs are fundamentally different from previously reported types. Consequently, a thorough investigation of these topological solitons is both necessary and valuable. To assess the stability of SK-LNs under an external magnetic field, we applied a magnetic field perpendicular to the surface (see Fig. 2(a–d)). As the field strength increased from 0 T to 0.5 T, the magnetic vortexes deformed from highly symmetric circles into ellipses. They exhibited strong topological protection, maintaining their structure up to a field strength of 0.7 T, which exceeds the critical field strength of most reported magnetic skyrmions. At 0.8 T, the vortexes were fully suppressed and transformed into helimagnetic strips. Remarkably, the skyrmions spontaneously regenerated once the external field was removed, demonstrating their exceptional non-volatile properties (see Fig. 2(e) and Extended Data Fig. 2 and 3).

To evaluate the impact of temperature on SK-LNs, we performed high-temperature LTEM experiments to assess their stability. Fig. 2(f–i, I) present under-focused LTEM images of the skyrmions at temperatures of 300 K (room temperature), 433 K, 593 K, and 673 K (the maximum testing temperature for the LTEM device). The results demonstrate that the skyrmions maintain a remarkably stable vortex structure even at the elevated temperature of 673 K, surpassing the stability observed in previous studies and indicating the exceptional high-temperature stability of SK-LNs. Given the ferroelectric nature of LN and the presence of DMI, we also investigated potential ferroelectric-related characteristics of the skyrmions. To distinguish these characteristics from magnetic signals, we adjusted the LTEM focal length to 0 µm (on-focused LTEM). The on-focused LTEM images (Fig. 2(f–i, II)) reveal a new grainy contrast in the center of the skyrmions at high temperatures, with domains displaying alternating dark and light stripes. This contrast suggests the presence of polar nanoregions (PNRs) within the central region of the skyrmions, indicating significant ferroelectric polarization[32,33]. Fig. 2(j) shows both under-focused and on-focused LTEM images of the LN film at 673 K, in a narrow field of view of 200 nm². The skyrmion remains stable while the PNR signal is observed. The size of the ferroelectric domain strips at the center of the skyrmion is approximately $6 \times 20$ nm², and the overall PNR structure exhibits a subcircular shape with a diameter of approximately 20 nm, highlighting the localized nature of the ferroelectricity within these skyrmions.

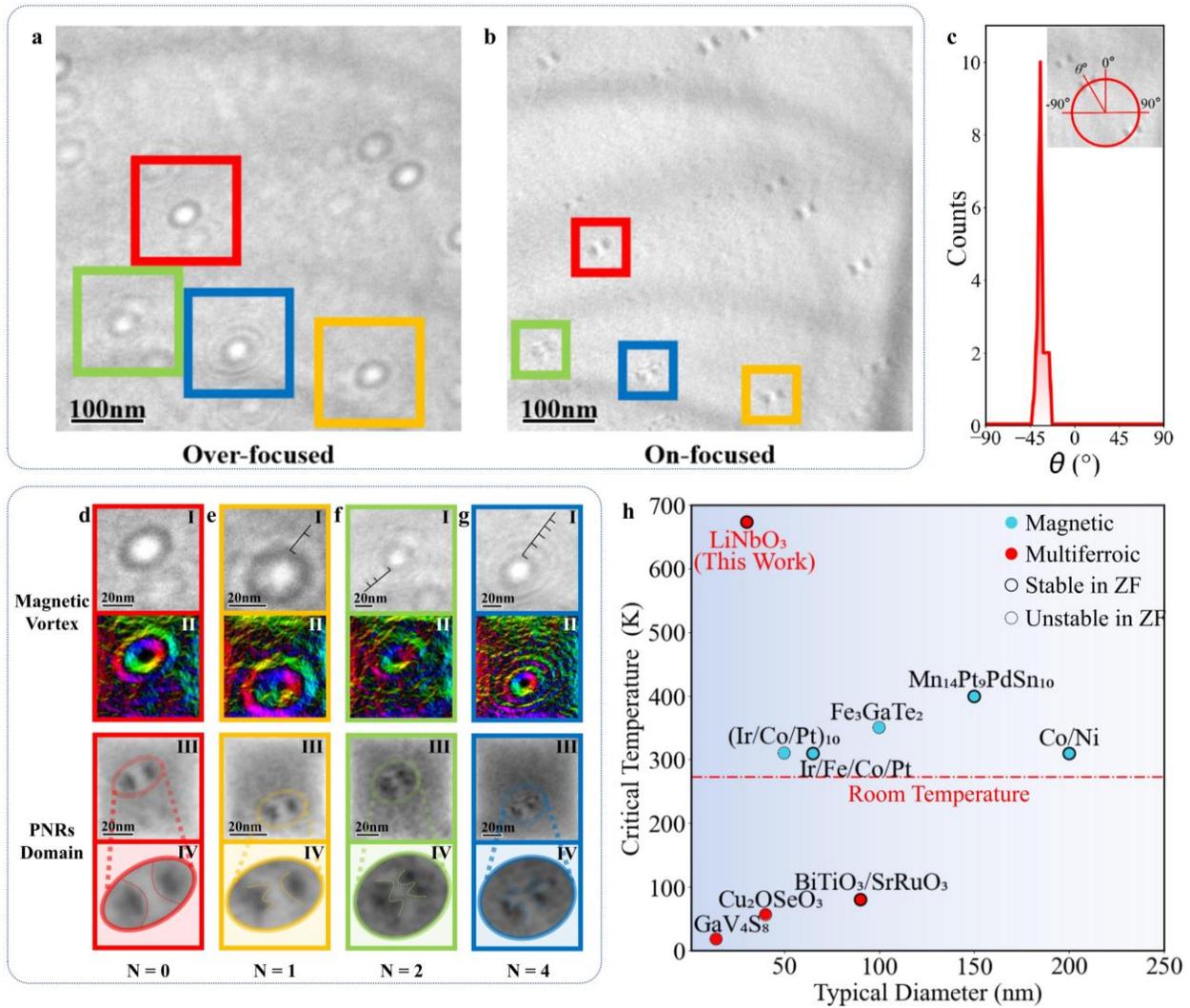

**Fig. 3 | PNR morphology with different types of skyrmions**. (**a,b**) Over-focused and on-focused images within an expansive field of vision, in which the skyrmions with specific ring number are indicated by blocks. (**c**) Statistical results of the PNR domain wall orientation. (I)

Over-focused LTEM images showing individual skyrmions with (**d**) N = 0, (**e**) N = 1, (**f**) N = 2, and (**g**) N = 4; the line segments are marked on each ring. (II) Corresponding magnetic induction maps. (III) Corresponding PNR maps. (IV) Corresponding PNR maps in a large view. The ferroelectric domain boundaries are delineated by dashed lines to facilitate resolution. (**h**) Statistical diagram illustrating the diameter and the maximum stable temperature for all multiferroic skyrmions reported and typical room temperature magnetic skyrmions. ZF: zero external magnetic field.

The over-focused and on-focused LTEM images of the LN film, captured at room temperature after cooling from 673 K, are shown in Fig. 3(a) and (b). These images demonstrate that the SK-LNs and their PNRs remain stable even without the high temperature. The images provide insight into the correlation between the ferroelectric domains (PNRs) and the magnetic vortices (SK-LNs). Firstly, PNRs in SK-LNs are highly localized within the magnetic vortex region, with no discernible PNR structures observed outside this domain. This suggests that the PNRs in SK-LNs are not solely a consequence of the intrinsic ferroelectricity of LN but rather a result of the ferroelectric response induced by the non-collinear magnetic structure, similar to the mechanism of magnetically-induced ferroelectricity. Although there is a possibility that LN's ferroelectricity might passively respond to the magnetic vortex through magnetoelectric coupling, this will be addressed and clarified in the following section. Additionally, the orientation of the domain walls of PNRs in SK-LNs was analyzed, revealing a high degree of consistency (Fig. 3(c)). Given the high symmetry of the magnetic vortices, this consistency can be attributed to the ferroelectric anisotropy of LN. Therefore, the observed PNRs likely result from the interaction between the intrinsic ferroelectric properties of SK-LNs and the underlying characteristics of LN. Another significant observation is that in anisotropic helimagnetic strips, the internal PNRs align with the orientation of the magnetic structure (see Extended Data Fig. 4). This alignment suggests that the anisotropy of the ferroelectricity induced by the magnetic structure disrupts the fixed orientation of the PNRs, which is influenced by the ferroelectric anisotropy of LN.

Furthermore, we identified several unique skyrmions with a multi-ring structure, resembling Hopfions and skyrmion bundles but featuring an increased number of rings[34,35]. Significant variations in the textures of the PNRs within these skyrmions were observed. These structures are illustrated in Fig. 3(d–g), where $N$ denotes the number of outer rings. Dotted lines mark the positions of the PNR domain walls to facilitate differentiation. For skyrmions with $N = 0$ and $N = 1$, the ferroelectric domains at the vortex center display alternating dark and light stripes, with the domain walls consistently oriented as previously described. For $N = 2$, the bright fringe at the center of the ferroelectric domain extends outward, causing the dark fringes on both sides to split, resulting in a point-like configuration. For $N = 4$, the central bright stripe protrudes outward and assumes a cross-like shape. These observations suggest that alterations in the ferroelectric morphology are linked to changes in the magnetic vortex structures of multiferroic skyrmions. As $N$ increases, the impact of LN's ferroelectric anisotropy on PNRs decreases, and the orientation of the domain walls becomes less constrained by the fixed direction typically associated with magnetoelectric coupling.

Comparing all reported multiferroic skyrmions (excluding stress-induced multiferroic heterojunctions[36,37], as the skyrmions in these systems do not exhibit ferroelectricity) and typical room-temperature magnetic skyrmions[38-45], as illustrated in Fig. 3(h), SK-LNs exhibit significant advantages in terms of stability at room temperature, reduced particle size, and magnetoelectric coupling. These exceptional properties are attributed to the unique physical mechanism of the Dzyaloshinskii-Moriya Interaction (DMI), which warrants further investigation and exploration for potential applications.

## Pockels effect of the SK-LNs and LN system

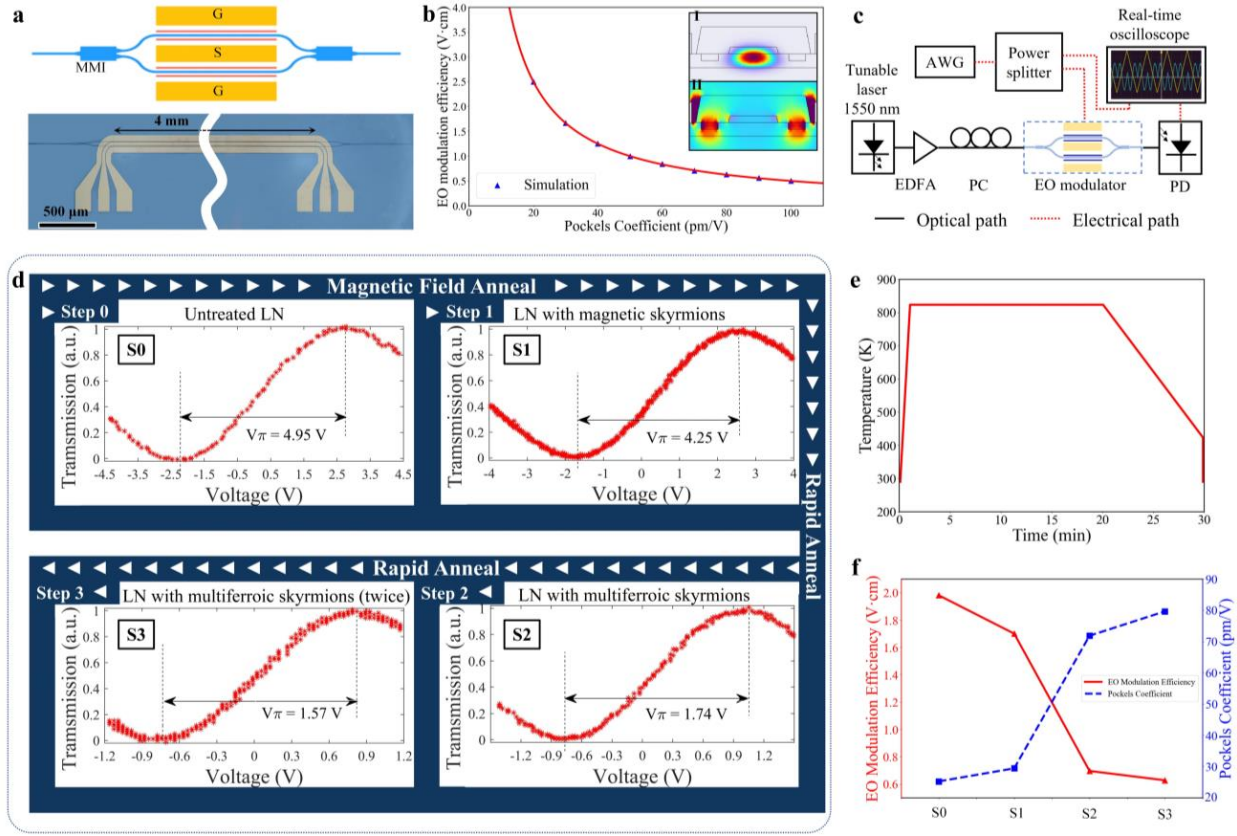

**Fig. 4** | **Test of EO modulator**. (**a**) Structure of the EO modulator under test; the microscope photograph showcases the fabricated device. The electrode positioned in the centre of the device is the signal electrode (S), and the remaining electrodes on both sides serve as ground electrodes (G). (**b**) The relationship between the Pockels coefficient and the half-wave voltage by simulation. (I) The intensity distribution of light in the waveguide (II) The distribution of electric field intensity across the cross-section of the modulation region of the device (**c**) Setup for the half-wave voltage measurement. AWG: Arbitrary signal generator. EDFA: Erbium-doped fibre

amplifier. PC: Polarisation controller. PD: High-speed photodetector. (**d**) Half-wave voltage test results of the EO modulator after undergoing multi-stage processing. The applied voltage frequency is 1 MHz, and the input wavelength is 1550 nm. (**e**) Process of the rapid annealing. (**f**) Statistical analysis of the EO modulation efficiency and the estimation of the Pockels coefficient.

After confirming the ferroelectric properties of SK-LNs, their potential impact was investigated, focusing on the linear EO effect, or Pockels effect, which is widely utilized in LN materials. We used an EO modulator based on the Mach-Zehnder structure to measure the Pockels coefficient by analyzing the relationship between the optical signal intensity variation and the applied electric field (as shown in Fig. 4(a)) [46]. The modulator was implemented on the X-cut TFLN platform with a 4 mm modulation length; additional design parameters are detailed in Extended Data Fig. 6 and Table 1. Simulations revealed the relationship between the Pockels coefficient and the half-wave voltage of the device, as depicted in Fig. 4(b) and Extended Data Fig. 7.

The test setup for the EO modulator is depicted in Fig. 4(c), with detailed descriptions provided in the Method. Results shown in Fig. 4(d) reveal that without skyrmion excitation, the modulator had a half-wave voltage (Vπ) of 4.95 V, corresponding to a modulation efficiency of 1.98 V·cm (S0). Following skyrmion formation via magnetic-field annealing (S1), Vπ decreased to 4.25 V. A subsequent rapid annealing (Fig. 4(e)), inducing PNRs within the magnetic vortexes (S2), further reduced Vπ to 1.74 V, significantly enhancing modulation efficiency. Additional annealing lowered Vπ to 1.57 V, approaching its threshold (S4). After all processes, modulation efficiency improved from 1.98 V·cm to 0.63 V·cm, with the Pockels coefficient exceeding 80 pm/V, representing a 2.2-fold increase (Fig. 4(f)). The evolution of the Pockels coefficient indicates that the ferroelectric properties of LN (S0 to S1) are not significantly enhanced by the magnetic vortex structure alone but are notably improved by the stimulated presence of PNRs (S1 to S2). This suggests that PNRs in skyrmions actively contribute to ferroelectric behavior, rather than merely responding passively to the magnetic vortexes. Consequently, SK-LNs exhibit inherent ferroelectric properties, distinguishing them from LN, and demonstrate a magnetoelectric coupling mechanism.

## Conclusion and outlook

In summary, our study confirms the presence of the DMI in LN through the excitation of magnetic skyrmions, validating the FE-induced-FM mechanism. We provided substantial evidence of magnetoelectric coupling in LN, including uniform domain wall orientation of PNRs and an enhanced Pockels effect, marking the first experimental confirmation of FE-induced-FM. This breakthrough has significant implications for future research on magnetoelectric coupling and multiferroic materials. Additionally, we achieved stable room-temperature multiferroic skyrmions, demonstrating potential for spintronics and advanced data storage technologies.

Furthermore, we discovered that DMI induced by FE-induced-FM can stimulate additional ferroelectricity, superimposing it onto the original ferroelectricity, similar to magnetically-induced ferroelectricity mechanisms. This process significantly enhances the ferroelectric properties of LN, resulting in a Pockels coefficient of up to 80 pm/V and enabling the realization of an electro-optic (EO) modulator with a remarkably low half-wave voltage of 0.63 V·cm. This EO modulator is pivotal for advancing large-scale optical computing and high-speed optical communication. Our research has markedly improved its performance using a straightforward process that is conducive to mass production, potentially leading to significant breakthroughs in optoelectronics.

Nevertheless, further research is required to address unresolved questions regarding this experiment. A key challenge is explaining why DMI in LN necessitates high temperature and an external magnetic field for excitation. We observed a low but non-zero probability of multiferroic skyrmions spontaneously occurring in untreated LN films (Extended Data Fig. 8), suggesting that magnetic field annealing and subsequent rapid annealing facilitate, rather than are essential for, skyrmion formation. Additionally, the complex magnetoelectric coupling mechanism in LN, particularly why DMI induced by FE-induced-FM also stimulates additional ferroelectricity, warrants further investigation. Understanding this mechanism will greatly contribute to the exploration of novel magnetoelectric coupling materials.